\documentclass[aps,prb,twocolumn,showpacs,superscriptaddress,floatfix]{revtex4-1}
\usepackage{dcolumn}
\usepackage{amsmath}
\usepackage{graphicx,epsfig,psfrag}
\usepackage{amssymb}
\usepackage{natbib}
\usepackage{url}
\usepackage[breaklinks=true]{hyperref}
\usepackage{mathtools}
\usepackage{subfigure}
\hypersetup{
        colorlinks=true,
}
\usepackage{bookmark}
\usepackage{epic}

\usepackage{txfonts}
\usepackage{color}
\newcommand{\pprI}{paper~I}

\begin{document}
\title{Time-Dependent Numerical Renormalization Group Method for Multiple Quenches: 
Application to General Pulses and Periodic Driving}
\author{H. T. M. Nghiem}
\affiliation
{Peter Gr\"{u}nberg Institut and Institute for Advanced Simulation, 
Research Centre J\"ulich, 52425 J\"ulich, Germany}
\author{T. A. Costi}
\affiliation
{Peter Gr\"{u}nberg Institut and Institute for Advanced Simulation, 
Research Centre J\"ulich, 52425 J\"ulich, Germany}
\begin{abstract}
The time-dependent numerical renormalization group method (TDNRG) [Anders {\em et al.,}, Phys. Rev. Lett. {\bf 95}, 196801 (2005)] 
was recently generalized to multiple quenches and arbitrary finite temperatures [Nghiem {\em et al.,} Phys. Rev. B {\bf 89}, 075118 (2014)] 
by using the full density matrix approach [Weichselbaum {\em et al.,} Phys. Rev. Lett. {\bf 99}, 076402 (2007)]. The formalism rests solely
 on the numerical renormalization group (NRG) approximation. In this paper, we numerically implement this formalism to study the response 
of a quantum impurity system to a general pulse and to periodic driving, in which a smooth pulse or a periodic train of pulses is 
approximated by a sufficient number of quenches. We show how the NRG approximation affects the trace of the projected density 
matrices and the continuity of the time-evolution of a local observable. We also investigate the long-time limit 
of a local observable upon switching from a given initial state to a given final state as a function of both the pulse shape and the 
switch-on time, finding that this limit is improved for smoother pulse shapes and longer switch-on times. 
This lends support to our earlier suggestion that the long-time limit of observables, following a quench 
between a given initial state and a given final state, can be improved 
by replacing a sudden large and instantaneous quench by a sequence of smaller ones acting over a finite time interval: longer 
switch-on times and smoother pulses, i.e., increased adiabaticity, favor relaxation of the system to its correct thermodynamic 
long-time limit. For the case of periodic driving, we compare the TDNRG results to the exact analytic ones for the non-interacting 
resonant level model, finding better agreement at short to intermediate time scales in the case of smoother driving fields. 
Finally, we demonstrate the validity of the multiple-quench TDNRG formalism 
for arbitrary temperatures by studying the time-evolution of the occupation number in the interacting Anderson impurity model in response to 
a periodic switching of the local level from the mixed valence to the Kondo regime at low, intermediate, and high temperatures.
\end{abstract}
\pacs{75.20.Hr, 71.27.+a, 72.15.Qm, 73.63.Kv}
% new: diamagnetism and paramagnetism, 75.20.Hr
% new: electronic conduction in metals and alloys, 72.15.Qm
% keep: 71.27.+a Strongly correlated electron systems; heavy fermions
%  73.21.La     Quantum dots
%  73.63.Kv     Quantum dots - electron. transport in nanoscale materials and structures
%Heat capacity of crystalline solids: 65.40.Ba
% entropy, thermodynamics, 05.70.-a
% removed: 73.20.-r Electron states at surfaces and interfaces
%removed: 71.30.+h      Metal-insulator transitions and other electronic transitions 

\date{\today}

\maketitle

\section{Introduction} 
\label{sec:introduction}

The time-evolution of strongly correlated systems in response to perturbations, such as quantum quenches, pulses, or 
periodic driving fields, is a topic of great current interest, with relevance to diverse fields, such as, for example,
pump-probe investigations of correlated materials \cite{Perfetti2006}, investigations of
coherent control and relaxation in solid-state qubits, \cite{Petta2005,Koppens2006} driven quantum tunneling, \cite{Grifoni1998} 
the determination of relaxation rates of excited spin states of magnetic adtoms via voltage pulses, \cite{Loth2010} 
and non-equilibrium effects in cold atom systems. \cite{Greiner2002,Will2010,Trotzky2012,Schneider2012}
A reliable understanding of the time-dependence of strongly correlated systems, in response to such perturbations, is an
important theoretical challenge. Open issues, include, for example, the description of thermalization following a 
quantum quench,\cite{Rigol2007,Moeckel2008,Polkovnikov2012} or the description of non-equilibrium states in response to stationary or time-dependent fields.
\cite{Rosch2003a,Anders2008,Schoeller2009}

For quantum impurity models, a number of techniques are available for studying time-dependent dynamics, including 
functional renormalization group,\cite{Metzner2012} generalized to time-dependent problems in Ref.~\onlinecite{Kennes2012a},
real-time renormalization group, \cite{Schoeller2009} time-dependent numerical renormalization group (TDNRG), 
\cite{Anders2005,Anders2006,Eidelstein2012,Guettge2013,Nghiem2014} perturbative scaling,\cite{Rosch2003a} 
Keldysh perturbation theory,\cite{Kamenev2011} real-time\cite{Saptsov2012} 
and renormalized perturbation theory,\cite{Bauer2005} flow-equation, \cite{Lobaskin2005,Wang2010} dual-fermion\cite{Jung2012}, 
slave-boson,\cite{Langreth1991} quantum Monte Carlo, \cite{Cohen2014,Gull2011,Weiss2008b,Muehlbacher2008} density matrix renormalization group (DMRG) 
for impurities embedded in one dimensional chains,
\cite{Daley2004,White2004,Schmitteckert2010} the time-dependent Gutzwiller approach,\cite{Schiro2010b,Lanata2012}, 
and $1/N$-expansion techniques\cite{Merino1998,Ratiani2009}. Applications of these to a number of
quantum impurity models have been made, including, to the interacting resonant level \cite{Andergassen2011,Karrasch2010,Kennes2012a,Kennes2012b}, 
and the Anderson impurity model\cite{Meir1993,Shao1994a,Goker2007,Oguri2001,Bauer2005,Munoz2013,Ratiani2009,Saptsov2012,Nghiem2014,Schiro2012}. 

In this paper, we shall be concerned with the TDNRG method. The underlying numerical renormalization group (NRG) method 
\cite{Wilson1975,KWW1980a,Gonzalez-Buxton1998,Bulla2008} 
has proven to be one of the most powerful and accurate methods for dealing with equilibrium properties of
strongly correlated quantum impurity systems, yielding essentially exact results.\cite{Hewson1997,Bulla2008,Merker2012b} 
Despite this, its application to time-dependent phenomena has revealed
a number of problems, such as difficulty in obtaining exactly the long-time limit of observables following a quantum quench, \cite{Rosch2012,Nghiem2014} or, difficulties in describing non-equilibrium steady states and non-equilibrium spectral functions \cite{Anders2008}.  These problems, 
together with techniques for extending the NRG to more complex  multi-channel models \cite{Campo2005,Ferrero2007,Costi2009,Hanl2013,Mitchell2014} 
are currently under active investigation. 

\begin{figure}[ht]
\centering
\includegraphics[width=0.4\textwidth]{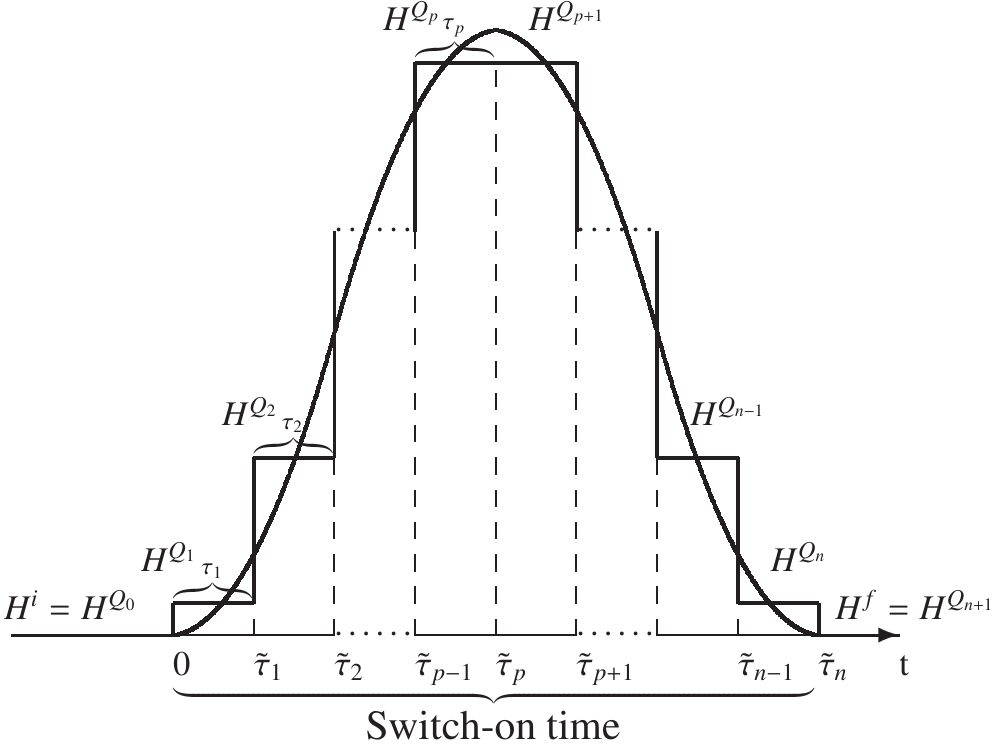}
\caption
{
  A system driven from an initial to a final state via a sequence of quantum quenches at times $\tilde{\tau}_0=0, \tilde{\tau}_1,\dots,\tilde{\tau}_n$
  with evolution according to $H^{Q_p}$ in the time step $\tilde{\tau}_p>t\geq\tilde{\tau}_{p-1}$. 
  Such a sequence of multiple quantum quenches could also be used to describe periodic switching, i.e, a periodic train of pulses, or, to 
  approximate any general continuous pulse (e.g., as indicated by the smooth solid line). For the case shown here, where initial and final states are
the same, the switch-on time corresponds to the pulse duration.
}
\label{fig:generic}
\end{figure}

In a previous paper (Ref.~\onlinecite{Nghiem2014}, henceforth referred to as \pprI), we presented a generalization of the time-dependent 
numerical renormalization group method (TDNRG) for single quantum quenches to finite temperatures within the full density matrix (FDM)
approach \cite{Weichselbaum2007}. The results of this finite temperature generalization of the TDNRG approach were illustrated 
by application to the Anderson impurity model. In addition, in \pprI{} we also generalized the finite-temperature formalism for the single quench case 
to multiple quantum quenches. Multiple quantum quenches can be used to describe general continuous pulses and periodic switching by a suitable discretization of the time domain as illustrated in  Fig.~\ref{fig:generic}. While the formalism for the multiple-quench case is 
considerably more complicated than that for the single-quench case, we showed in \pprI{} that it is nevertheless numerically feasible. In particular, 
we showed that the computational time should scale approximately linearly with the number $n_{\rm quench}$ of quenches.
In this paper, we implement this approach numerically for the Anderson impurity model and its non-interacting counterpart, the resonant level model 
(RLM), and present results for two interesting situations, (i), general pulses 
acting over a finite time interval, the so called called switch-on time $\tilde{\tau}_n$ [see Fig.~\ref{fig:generic}], and, (ii), periodic driving where a system 
parameter, such as the local level position, is modulated periodically in time. Periodic driving has also been studied for the interacting resonant level model in 
Ref.~\cite{Eidelstein2012}, by using a hybrid TDNRG method, combining the TDNRG at short times with the Chebyshev expansion technique\cite{Weisse2006} 
for longer times. In contrast to the TDNRG approach used in 
Ref.~\cite{Eidelstein2012}, which involved additional approximations beyond the NRG approximation, our TDNRG formalism rests 
solely on the latter approximation (see Sec.~\ref{subsec:formalism} and \pprI{} for details).

The paper is organized as follows. In Sec.~\ref{sec:prelims}, we outline the model and the notation for describing multiple-quenches, 
provide a brief description of the NRG, the complete basis set and
the FDM, and recapitulate the multiple-quench TDNRG formalism from \pprI{}. For 
full details of the derivation of the multiple-quench TDNRG, we refer the reader to the previous publication. 
In Sec.~\ref{sec:limits} we discuss exact results and limiting cases, the conservation 
of the trace of the projected density matrices in each time interval,
and the continuity of observables at the boundaries of these time intervals, 
and how these are affected by the use of the NRG approximation. We argue in Sec.~\ref{sec:limits} that the NRG approximation 
introduces a cumulative error in the trace of the projected density matrices after all but the
first quantum quench, and discuss the size of this finite time error as well as its influence on the continuity of observables. 
In Sec.~\ref{sec:numerical},  we present our numerical results of the multiple-quench formalism, applied to general pulses for the 
Anderson impurity model (Sec.~\ref{subsec:general-pulses}) and to periodic driving for the RLM and Anderson impurity models (Sec.~\ref{subsec:periodic-driving}).
In the former, we analyze the error in the long-time limit of observables, 
both as a function of the switch-on time for a fixed pulse shape (a linear ramp) 
, and also its dependence on different pulse shapes, such as linear, trigonometric, and logistic, for a fixed switch-on time. 
In the latter, periodic driving is investigated for square and triangular pulses, comparing with analytical continuum results for the case of the 
RLM, which allows us to demonstrate the accuracy and limitations of the method. 
In addition, periodic driving is applied
to the strong correlation limit of the Anderson impurity model in a wide range of temperatures, thereby demonstrating the application of the
formalism to finite temperatures. Conclusions and an outlook are given in Sec. \ref{sec:summary}.

\section{Preliminaries}
\label{sec:prelims}
\subsection{Model, multiple quenches and time-evolution}
\label{subsec:models}
We shall apply the TDNRG method for multiple quenches and general pulses to the model defined by
\begin{align}
H&= H_{\rm imp}+ H_{\rm bath} + H_{\rm int},\\
H_{\rm imp}&= \sum_{\sigma}\varepsilon_d(t)n_{d\sigma}+U(t)n_{d\uparrow}n_{d\downarrow},\\
H_{\rm bath}&=\sum_{k\sigma}\epsilon_k c^{\dagger}_{k\sigma}c_{k\sigma},\\
H_{\rm int}&=\sum_{k\sigma}V(t)c^{\dagger}_{k\sigma}d_{\sigma}+h.c.
\end{align}
Here, $n_{d\sigma}=d^{\dagger}_{\sigma}d_{\sigma}$ is the number operator for electrons with spin $\sigma$ in a local level with energy $\varepsilon_d(t)$. The Coulomb repulsion between two electrons in the local level is $U(t)$, $\epsilon_k$ is the kinetic energy of the conduction electrons with wavenumber $k$, and $V(t)$ is the hybridization matrix element of the local d-state with the conduction states. For  $U(t)\neq 0$, this model corresponds to the Anderson impurity model which for $U(t)=0$ 
reduces to the non-interacting resonant level model (RLM).

In this paper, we consider a switching protocol in which only the local level position is allowed to have a time-dependence $\varepsilon_d(t)$, with $V(t)=V$ and
$U(t)=U$ being kept constant. 

Both the Anderson model and its non-interacting counterpart, the RLM, are characterized by the bare energy scales $\varepsilon_d$ and the hybridization strength $\Gamma=\pi\rho V^2$, where $\rho=1/W$ is the density of state for a flat band of width $W=2D=2$, and $D=1$ is the half-bandwidth. In the case of the Anderson model in the strong correlation limit, $U\gg\Gamma$, and for $-\varepsilon_{d}\gg\Gamma$, an additional  low energy scale emerges, the Kondo scale $T_{\rm K}=\sqrt{U\Gamma/2}e^{\pi\varepsilon_d(\varepsilon_d+U)/2\Gamma U}$. We shall express the temperature in terms of $T/{\Gamma}$ for the RLM calculations, and in terms of $T/{T_{\rm K}}$ for the Anderson impurity model calculations. In plotting the time-dependence of local observables, we shall use a time-axis variable $t\Gamma$ in all cases, 
i.e., time is measured in units of $\hbar/\Gamma$ with the Planck constant set to unity $\hbar=1$.

For a system driven through a set of quenches, as in Fig.~\ref{fig:generic}, the time-evolved density matrix at a general time in the interval $\tilde{\tau}_{p+1}>t\geq\tilde{\tau}_p$ is given by
\begin{widetext}
\begin{align}
\rho(t)=e^{-iH^{Q_{p+1}}(t-\tilde{\tau}_p)}e^{-iH^{Q_p}{\tau}_p}...e^{-iH^{Q_1}{\tau}_1}\rho e^{iH^{Q_1}{\tau}_1}...e^{iH^{Q_p}{\tau}_p}e^{iH^{Q_{p+1}}(t-\tilde{\tau}_p)},\label{eq:rhot}
\end{align}
\end{widetext}
in which $H^{Q_{p}},p=1,\dots,n$ are intermediate quench Hamiltonians, acting during time intervals of length $\tau_{p},p$, that 
determine the time-evolution at intermediate times and $H^{Q_0}=H^{i}$ and $H^{Q_{n+1}}=H^f$ are the initial and final state Hamiltonians, respectively 
(see Fig.~\ref{fig:generic}), and $\rho$ is the initial density matrix of the system at time $t<\tilde{\tau}_0=0$ (to be specified in the next section). 
The time-evolution of a local observable $\hat{O}$ at $\tilde{\tau}_{p+1}>t\geq\tilde{\tau}_p$ is then given by 

\begin{align}
O(t)={\rm Tr}[\rho(t)\hat{O}].\label{eq:Ot}
\end{align}

%
%figure 1 was here, now moved
%

\subsection{NRG, complete basis set, and FDM}
\label{subsec:nrg+cbs}
The $n$ intermediate quench Hamiltonians $H^{Q_{p}},p=1,\dots,n$, together with initial, $H^{i}=H^{Q_0}$, and final, $H^{f}=H^{Q_{n+1}}$, state Hamiltonians are
iteratively diagonalized in the usual way within the NRG method \cite{Wilson1975,KWW1980a,Bulla2008}, yielding eigenstates and eigenvalues 
for a sequence of truncated Hamiltonians $H_{m}^{Q_p}, m=1,2,\dots$, which approximate the spectra of $H^{Q_p}$, on successively decreasing energy
scales $\omega_{m}\sim \Lambda^{-m/2}$. The discretization parameter $\Lambda>1$ is
required to achieve a separation of energy scales in $H^{Q_p}$, such that an iterative diagonalization scheme remains a controlled numerical procedure. 
This procedure is performed up to a maximum iteration $m=N$ (``the longest Wilson chain'').  At each $m$, the states generated, denoted $|qm\rangle_{Q_{p}}$,
are partitioned into the lowest energy retained states, denoted $|km\rangle_{Q_p}$,  and the high energy eliminated (or discarded) states, $| lm\rangle_{Q_p}$. 
In order to avoid an exponential increase
in the dimension of the Hilbert space, only the former are used to set up and diagonalize the Hamiltonian for the next iteration $m+1$. The eliminated states, 
while not used in the iterative NRG procedure, are nevertheless crucial as they are used to set up a complete basis set with which the expressions for the time dependent dynamics 
are evaluated.\cite{Anders2005} This complete basis set is defined by the product states $|lem\rangle_{Q_p}=|lm\rangle_{Q_p} |e\rangle, m=m_{0},\dots,N$, where $m_{0}$ is the first iteration at which truncation occurs,
and $|e\rangle=|\alpha_{m+1}\rangle |\alpha_{m+2}\rangle\dots|\alpha_{N}\rangle$ are environment states at iteration $m$ such that the product states $| lem\rangle_{Q_p}$, for each $m=m_{0},m_{0}+1,\dots,N$, 
reside in the same Fock space (that of the largest system diagonalized, $m=N$). The $\alpha_m$ represent the configurations of site $m$ in a linear chain representation of the quantum impurity system 
(e.g. the four states $0$, $\uparrow$, $\downarrow$ and $\uparrow\downarrow$ at site $m$ for a single channel Anderson model) and $"e"$ in $|lem\rangle_{Q_p}$ denotes the
collection $e=\{\alpha_{m+1}...\alpha_N\}$.  For each quench Hamiltonian $H^{Q_p}$, completeness relations may be defined \cite{Anders2005,Anders2006}
\begin{align}
&\sum_{m=m_0}^N \sum_{le} | lem\rangle_{Q_p}{_{Q_p}}\langle lem|=1,\label{unity-decomposition}
\end{align}
where for $m=N$ all states are counted as discarded (i.e. there are no kept states at iteration $m=N$).

By using the complete basis set for the initial Hamiltonian $H^{Q_0}=H^i$, we can construct an initial state density matrix $\rho$, entering Eq.~(\ref{eq:rhot}), and which is valid at any temperature, the FDM, \cite{Weichselbaum2007,Weichselbaum2012} 
\begin{align}
&\rho=\sum_{m=m_0}^N w_m \tilde{\rho}_m,\label{eq:fdm-initial-state}\\
&\tilde{\rho}_m=\sum_{le}|lem\rangle{_i} \frac{e^{-\beta E_l^m}}{\tilde{Z}_m}{_i}\langle lem|,\label{eq:rhom}
\end{align}
which includes all discarded states of $H^i$ from all shells. For later use, we note that, 
(a), ${\rm Tr}\left[\tilde{\rho}_{m}\right]={\rm Tr}\left[\rho\right]=1$ implies that $\sum_{m=m_0}^{N}w_m =1$, and, 
(b),  ${\rm Tr}\left[\tilde{\rho}_m\right]=1$ implies that $1=\sum_{le}\frac{e^{-\beta E_{l}^m}}{\tilde{Z}_{m}}=\sum_{l}d^{N-m}\frac{e^{-\beta E_{l}^m}}{\tilde{Z}_{m}}=d^{N-m}\frac{Z_{m}}{
\tilde{Z}_{m}}$ where $Z_{m}=\sum_{l}e^{-\beta E_{l}^{m}}$, i.e., $\tilde{Z}_m = d^{N-m}Z_m$, and $d$ is the degeneracy of the Wilson site $\alpha_m$ 
(with $d=4$ for the Anderson impurity model of this paper). \cite{Weichselbaum2007,Costi2010}

\subsection{Multiple-quench formalism}
\label{subsec:formalism}
With the above notation and background information, we recall the important equations in our multiple-quench TDNRG formalism given in \pprI{}. 
A system driven through a sequence of quenches,  as  in Fig.~\ref{fig:generic}, results in the following time-evolution for a local observable $\hat{O}$ 
at $\tilde{\tau}_{p+1}>t\geq\tilde{\tau}_p$ 
\begin{align}
O(t)=\sum_{mrs}^{\notin KK'}\rho^{i\to Q_{p+1}}_{rs}(m,\tilde{\tau}_p) e^{-i(E^m_r-E^m_s)(t-\tilde{\tau}_p)} O^m_{sr}\label{eq:Otime},
\end{align}
\begin{align}
 &\text{with}\quad\rho^{i\to Q_{p+1}}_{rs}(m,\tilde{\tau}_p)\nonumber\\
&= \sum_e{_{Q_{p+1}}}\langle rem|e^{-iH^{Q_p}{\tau}_p}\dots e^{-iH^{Q_1}{\tau}_1}\rho e^{iH^{Q_1}{\tau}_1}\dots e^{iH^{Q_p}{\tau}_p}| sem\rangle{_{Q_{p+1}}},\nonumber
\end{align}
and $O_{sr}^{m}={_{Q_{p+1}}}\langle sm|\hat{O}|rm\rangle_{Q_{p+1}}$ the matrix elements of $\hat{O}$.
Substituting the FDM of the initial state from Eqs.~({\ref{eq:fdm-initial-state}}) and (\ref{eq:rhom}) 
into the above projected density matrix, we have
\begin{align}
&\rho^{i\to Q_{p+1}}_{rs}(m,\tilde{\tau}_p)=\sum_{m_1l_1e_1e}{_{Q_{p+1}}}\langle rem|e^{-iH^{Q_p}{\tau}_p}\dots e^{-iH^{Q_1}{\tau}_1}|l_1e_1m_1\rangle{_i}\nonumber\\
&\hspace{3em}\times w_{m_1} \frac{e^{-\beta E_{l_1}^{m_1}}}{\tilde{Z}_{m_1}}{_i}\langle l_1e_1m_1|e^{iH^{Q_1}{\tau}_1}\dots e^{iH^{Q_p}{\tau}_p}| sem\rangle_{Q_{p+1}}\label{eq:rhomultiquenches}.
\end{align}
We decompose $\rho^{i\to Q_{p+1}}_{rs}(m,\tilde{\tau}_p)$ into three terms,
\begin{align}
\rho^{i\to Q_{p+1}}_{rs}(m,\tilde{\tau}_p)=&{\tilde{\rho}}^{++}_{rs}(m,\tilde{\tau}_p)+\rho^{0}_{rs}(m,\tilde{\tau}_p)+\rho^{--}_{rs}(m,\tilde{\tau}_p),\label{eq:Totalrhomultiquenches}
\end{align}
corresponding to the $m_1>m$, $m_1=m$, and $m_1<m$ contributions in Eq.~(\ref{eq:rhomultiquenches}) and following the same notation as in \pprI{}.
Explicitly written out, these are given by
\begin{align}
{\tilde{\rho}}^{++}_{rs}(m,\tilde{\tau}_p)=&\sum_{kk'}\mathcal{S}^m_{r_{Q_{p+1}}k_i}(-\tilde{\tau}_p)R^m_{\rm red}(k,k')\mathcal{S}^m_{k'_is_{Q_{p+1}}}(\tilde{\tau}_p)\label{eq:rho++multiple}\\
\rho^{0}_{rs}(m,\tilde{\tau}_p)=&\sum_{l}\mathcal{S}^m_{r_{Q_{p+1}}l_i}(-\tilde{\tau}_p)w_{m} \frac{e^{-\beta E_{l}^{m}}}{{Z}_{m}}\mathcal{S}^m_{l_is_{Q_{p+1}}}(\tilde{\tau}_p)\label{eq:rho0multiple}\\
\rho^{--}_{rs}(m,\tilde{\tau}_p)=&\frac{1}{d}\sum_{kk'\alpha_{m}}A^{\alpha_{m}\dagger}_{rk}\bigg\{\rho^0_{kk'}(m-1,\tilde{\tau}_p)+\rho^{--}_{kk'}(m-1,\tilde{\tau}_p)\bigg\}A^{\alpha_{m}}_{k's}\nonumber\\
& \text{with\quad } \rho^{--}_{rs}(m_0,\tilde{\tau}_p)=0\label{eq:rho--multiple},
\end{align}
in which $R^m_{\rm red}(k,k')$ is the full reduced density matrix of the initial state \cite{Weichselbaum2007,Costi2010}, and can be calculated recursively. \cite{Nghiem2014} 
The transformation matrix
$A^{\alpha_{m}}_{k's}$ entering above, relates eigenstates $|sm\rangle_{Q_p}$ of $H_{m}^{Q_p}$ to product states
$|k'm-1\rangle|\alpha_{m}\rangle_{Q_p}$,  i.e.,
\begin{equation}
|sm\rangle_{Q_p} =
\sum_{k'\alpha_{m}}A^{\alpha_{m}}_{k's}|k'm-1\rangle|\alpha_{m}\rangle.\label{eq:transformation-matrix}
\end{equation}
The generalized overlap matrix elements $\mathcal{S}^{m}_{r_is_{Q_{p+1}}}(\tilde{\tau}_p)$ appearing in Eqs.~(\ref{eq:rho++multiple}-\ref{eq:rho0multiple})
are diagonal in the environment variables \cite{Nghiem2014}, 
\begin{align}
\mathcal{S}^{m}_{r_is_{Q_{p+1}}}(\tilde{\tau}_p)\times\delta_{ee'}={_i}\langle rem|e^{iH^{Q_1}{\tau}_1}\dots e^{iH^{Q_p}{\tau}_p}|se'm\rangle{_{Q_{p+1}}}\label{eq:overlapTime}.
\end{align}
and assume the values $\mathcal{S}^m_{r_is_{Q_{1}}}(\tilde{\tau}_0)=S^m_{r_is_{Q_{1}}}$ at $\tilde{\tau}_0=0$, 
with $S^m_{r_is_{Q_{1}}}$ being the ordinary overlap matrix elements between eigenstates of $H^{Q_0}=H^i$ and $H^{Q_1}$. 
\footnote{The factor $e^{-iH^{Q_p}{\tau}_p}\dots e^{-iH^{Q_1}{\tau}_1}$ is the time-evolution operator
at time $\tilde{\tau}_{p}=\sum_{i=1}^{p}\tau_i$ following $p$ intermediate quantum quenches  described by $H^{Q_1},\dots,H^{Q_{p}}$. 
Hence, $\protect{{_{Q_{p+1}}}\langle rem|e^{-iH^{Q_p}{\tau}_p}\dots e^{-iH^{Q_1}{\tau}_1}|se'm\rangle{_i}}$
is the matrix element of this time-evolution operator between the initial states of $H^{i}$ and the states of the quench 
Hamiltonian $H^{Q_{p+1}}$. These generalized overlap matrix elements reduce to
ordinary overlap matrix elements only at $\tilde{\tau}_0=0$.
}

Similarly to the projected density matrix, the generalized overlap matrix elements $\mathcal{S}^m_{s_ir_{Q_{p+1}}}(\tilde{\tau}_p)$ can be decomposed into three terms and calculated recursively as follows:
\begin{align}
&\mathcal{S}^m_{r_is_{Q_{p+1}}}(\tilde{\tau}_p)=\mathcal{S}^{m++}_{r_is_{Q_{p+1}}}(\tilde{\tau}_p)+\mathcal{S}^{m0}_{r_is_{Q_{p+1}}}(\tilde{\tau}_p)+\mathcal{S}^{m--}_{r_is_{Q_{p+1}}}(\tilde{\tau}_p),\label{eq:totalStau}
\end{align}
\begin{align}
&\mathcal{S}^{m++}_{r_is_{Q_{p+1}}}(\tilde{\tau}_p)=\sum_{k}\mathcal{S}^m_{r_ik_{Q_p}}(\tilde{\tau}_{p-1})e^{iE^m_k{\tau}_p}S^m_{k_{Q_p}s_{Q_{p+1}}},\label{eq:totalStau++}\\
&\mathcal{S}^{m0}_{r_is_{Q_{p+1}}}(\tilde{\tau}_p)=\sum_{l}\mathcal{S}^m_{r_il_{Q_p}}(\tilde{\tau}_{p-1})e^{iE^m_l{\tau}_p}S^m_{l_{Q_p}s_{Q_{p+1}}},\label{eq:totalStau0}
\\
&\mathcal{S}^{m--}_{r_is_{Q_{p+1}}}(\tilde{\tau}_p)
=\sum_{\alpha_m}\sum_{kk'}A^{\alpha_m\dagger}_{rk}\Big[\mathcal{S}^{(m-1)0}_{k_ik'_{Q_{p+1}}}(\tilde{\tau}_{p})+\mathcal{S}^{(m-1)--}_{k_ik'_{Q_{p+1}}}(\tilde{\tau}_{p})\Big]A^{\alpha_m}_{k's},\label{eq:totalStau--}\nonumber\\
&\hspace{6em}\text{with\quad } \mathcal{S}^{m_0--}_{r_is_{Q_{p+1}}}(\tilde{\tau}_p)=0.\end{align}
For detailed proofs of these equations, we refer the reader to our previous paper. In reducing the generalized overlap matrix elements from the general expression in Eq.~(\ref{eq:overlapTime}) to the expressions in terms of the components in Eqs.~(\ref{eq:totalStau++})-(\ref{eq:totalStau--}), the NRG approximation is adopted in the time-evolution factors $e^{iH^{Q_p}{\tau}_p}$, i.e, we use that $H|qem\rangle\approx H^m|qem\rangle=E^m_q|qem\rangle$ and $e^{iHt}|qem\rangle\approx e^{iH^m t}|qem\rangle=e^{iE^m_qt}|qem\rangle$. Therefore, all the generalized overlap matrix elements are subject to an error coming from the NRG approximation, except for $\mathcal{S}^m_{r_is_{Q_{1}}}(\tilde{\tau}_0)=S^m_{r_is_{Q_{1}}}$ which involves no time-evolution factors. Furthermore, the generalized overlap matrix elements at $\tilde{\tau}_p$ depend recursively on those at $\tilde{\tau}_{p-1}$, Eqs.~(\ref{eq:totalStau++}) and (\ref{eq:totalStau0}). Hence the error in the generalized overlap matrix elements at $\tilde{\tau}_p$ is not just due to the NRG approximation at this step, but also accumulates from the error of overlap matrix elements at previous steps. Since the multiple-quench TDNRG formalism relies on 
the NRG approximation for evaluating the generalized overlap matrix elements, this approximation results in errors in the projected density matrices for $p>1$, which result also in errors in the time-evolution of a local observable, which we shall quantify in the next section on numerical results.

Nevertheless, our multiple-quench TDNRG formalism, is based solely on the NRG approximation, and explicitly includes all the components of the projected density matrix and generalized overlap matrix elements. As for the single quench case, discussed in our previous paper, the present multiple-quench TDNRG remains valid at arbitrary finite temperatures due to the use of the FDM in Eq.~(\ref{eq:rhomultiquenches}), see Sec.~\ref{subsubsec:temperature-dependence}. It differs from previous studies of multiple quenches within the hybrid TDNRG,\cite{Eidelstein2012} where some components, e.g., $\mathcal{S}^{m--}$ and $\rho^{--}$,  are neglected.

\section{Exact results and sources of error}
\label{sec:limits}
As for the single quench case\cite{Nghiem2014}, the TDNRG for multiple quenches obeys a number of exact results and fulfills some exact limits. 
We can use these  to check the numerical precision of the calculations as well as the accuracy of the multiple-quench TDNRG method. For example, 
in \pprI{} we showed that for the single quench case, the short-time limit of observables is exact, i.e., that $O(t\to 0^{+})=O_i$, 
with $O_i = {\rm Tr} [\rho \hat{O}]$ being the thermodynamic value in the initial state. This result remains true also for the multiple-quench TDNRG method : in both cases it relies
on the fact that the NRG approximation is inoperative in the limit $t\to 0^{+}$ 
(see \pprI{} for a formal proof of this result). In contrast to this, we find, as for the single quench case, that the long-time limit of 
observables, $O(t\to\infty)$, has a finite error,  which we shall discuss further in Sec.~\ref{subsec:general-pulses} in the context of general pulses. 
We next list the various limiting cases and exact results for the multiple-quench TDNRG.

First, if the switch-on time is set to zero, $\sum_n^{p=1} \tau_p = \tilde{\tau}_n=0$, the multiple-quench formalism reduces exactly to our formalism for the 
finite temperature single quench case in \pprI{}. In this case, each time interval $\tau_p = 0$ leads to $e^{iH^{Q_p}{\tau}_p}|rem\rangle_{Q_p}\equiv 1\dot |rem\rangle_{Q_p}$, 
and no error is incurred in these factors upon adopting the NRG approximation $H^{Q_p}|rem\rangle_{Q_p}\approx H_{m}^{Q_p}|rem\rangle_{Q_p}$. 
Therefore, from Eqs.~(\ref{eq:totalStau}-\ref{eq:totalStau--}) we have that
$\mathcal{S}^m_{r_is_{f}}(\tilde{\tau}_n=0)=S^m_{r_is_{f}}$, and the multiple-quench formalism recovers the single quench one exactly. 
Indeed, multiple-quench numerical calculations with $\tilde{\tau}_n=0$ yielded results within $10^{-10}$ of the corresponding single quench numerical 
calculation.

Second, when all quench sizes equal zero, i.e., when $H^{i}=H^{Q_0}=\dots = H^{Q_{n+1}}=H^{f}$, 
the expectation value of a local observable in Eq.~(\ref{eq:Otime}) is time-independent and exactly equals the equilibrium thermodynamic value $O(t)=O_{i}=O_{f}$. In this case, $S^m_{r_{Q_p}s_{Q_{p+1}}}=\delta_{rs}$ and the generalized overlap matrix elements in Eqs.~(\ref{eq:totalStau}-\ref{eq:totalStau--}) yield $\mathcal{S}^m_{r_is_{Q_{p+1}}}(\tilde{\tau}_p)=e^{iE^m_r\tilde{\tau}_p}\delta_{rs}$. Substituting this generalized overlap matrix element into Eqs.~(\ref{eq:rho++multiple}-\ref{eq:rho0multiple}), we see that only the $\rho^{0}_{rs}(m,\tilde{\tau}_p)$ component of the projected density matrix contributes in Eq.~(\ref{eq:Otime}) with the restriction $rs\notin KK'$. Even though, the NRG approximation appears in $\mathcal{S}^m_{r_is_{Q_{p+1}}}(\tilde{\tau}_p)$, it will be canceled by the complex conjugate $\mathcal{S}^m_{s_{Q_{p+1}}r_i}(-\tilde{\tau}_p)$ in Eq.~(\ref{eq:rho0multiple}). Eventually, $O(t)=\sum_{ml}w_{m} \frac{e^{-\beta E_{l}^{m}}}{{Z}_{m}}O_{ll}=O_i=O_f$. In this limiting case,  the multiple-quench numerical calculation agreed with the exact equilibrium results 
within an error of typically less than $10^{-10}$. Together with the first exact result above, this provides a useful test that the multiple-quench formalism is correctly implemented numerically.

Third, by setting the local observable in Eq.~(\ref{eq:Otime}) to be the identity operator, $\hat{O}=1$, we can show that the trace of the 
projected density matrices is preserved at each time step, i.e., ${\rm Tr}[\rho^{i\to Q_{p+1}}(\tilde{\tau}_p)]=1$, provided that the NRG approximation is not used
to evaluate the generalized overlap matrix elements in Eq.~(\ref{eq:overlapTime}). The proof of this may be found in Appendix~\ref{sec:trace-projected-density-matrix}.
In practice, as outlined at the end of  Sec.~\ref{sec:limits}, the evaluation of these generalized overlap matrix elements proceeds via the recursive expressions in 
Eqs.~(\ref{eq:totalStau}-\ref{eq:totalStau--}) which are obtained by making the NRG approximation. Hence, except for the first time interval $t\leq {\tilde{\tau}}_{1}$,
an accumulated error in the trace of the projected density matrices for $p>1$ will arise, which we shall quantify in more detail, 
numerically, below.

Finally, one can ask whether the continuity in the time-evolution of a local observable 
is guaranteed at the boundaries of each time step, i.e., whether
$O(t\to \tilde{\tau}^-_p)=O(t\to \tilde{\tau}^+_p)$. % = O(\tilde{\tau}_p)$. 
Here, again, we can prove that this holds within the multiple-quench TDNRG formalism presented above, provided that we do not use
the NRG approximation in evaluating the generalized overlap matrix elements in Eq.~(\ref{eq:overlapTime}), see Appendix~\ref{sec:continuity}.
However, in practice, this approximation is required to arrive at a feasible procedure for the evaluation of these matrix elements, such as the
recursion relations in Eqs.~(\ref{eq:totalStau}-\ref{eq:totalStau--}), which are obtained from Eq.~(\ref{eq:overlapTime}) by making use of only
the NRG approximation. Consequently, in actual numerical calculations, which use Eqs.~(\ref{eq:totalStau}-\ref{eq:totalStau--}), 
discontinuities in the time dependence of observables at the boundaries of the time intervals do arise and we shall discuss these 
in more detail below (see also the next section on numerical results, particularly Sec.~\ref{subsubsec:dependence-switch-on-time}).

\begin{figure}[ht]
  \centering
    \includegraphics[width=0.5\textwidth]{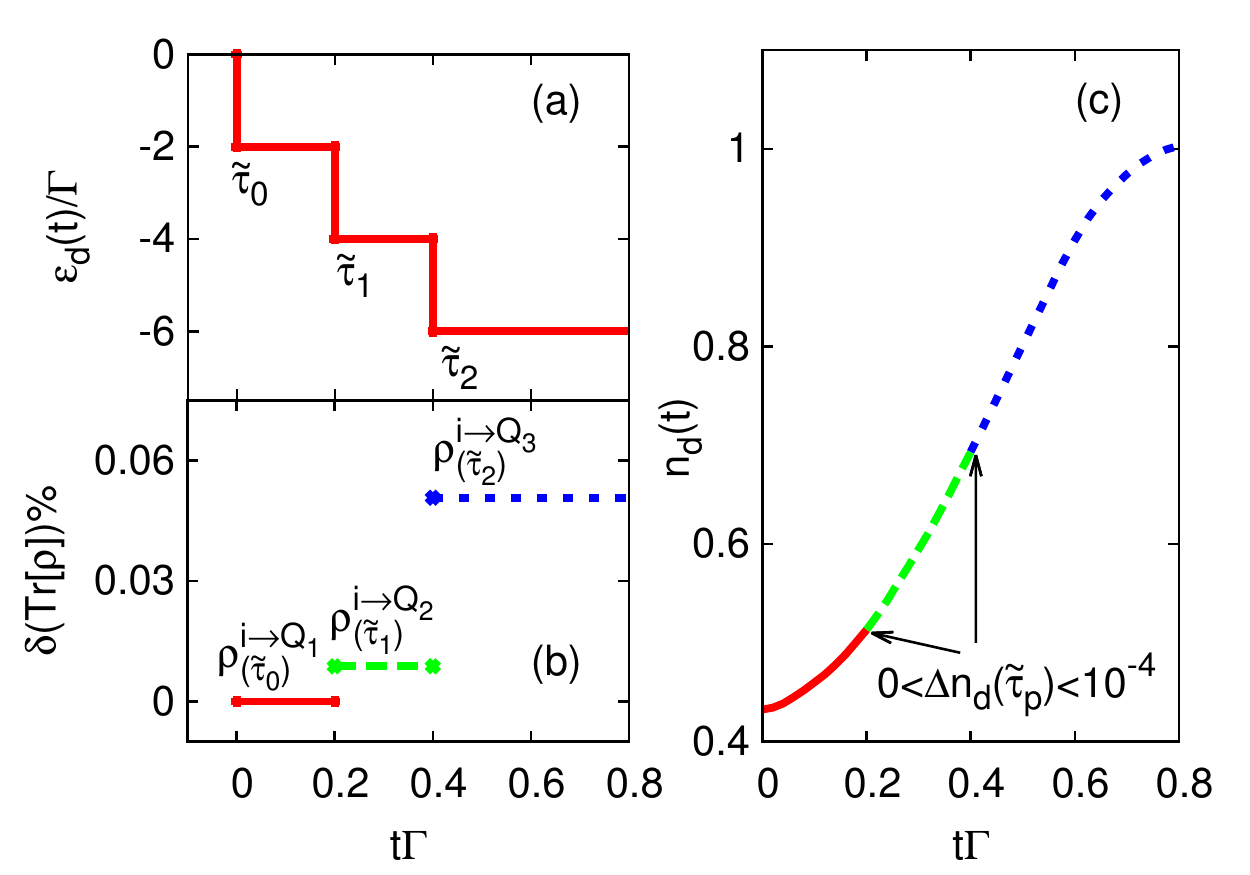}
  \caption
{
  (Color online) Application of the multiple-quench TDNRG to the Anderson impurity model, for $n_{\rm quench}=3$ 
quenches at low temperature $T/T_{\rm K}\approx 10^{-4}$ (essentially zero temperature). The sequence of quenches, shown in (a), 
switches the system from $\varepsilon_i=0$ (the mixed valence regime) through two states with local level position at $-2\Gamma$ and $-4\Gamma$, before 
eventually switching it to the final state at $\varepsilon_f=-6\Gamma$ (the symmetric Kondo regime).
(b) The deviation of ${\rm Tr}[\rho^{i\to Q_p}(\tilde{\tau}_{p-1})]$ from the expected value of $1$ at each time step. (c) The time-evolution of the
occupation number at each time step.
  The other parameters are $U=12\Gamma$, and $\Gamma=10^{-3}D$. $T_{\rm K}\approx 2.0\times 10^{-5}D$ is the Kondo temperature 
in the final state. 
  The calculations are for $\Lambda=4$, no $z$ averaging, and keeping states below $E_{\rm cut}=24$.
In this, and all subsequent numerical results, 
we do not use any damping\cite{Anders2006} in the exponential factors appearing in Eq.~({\ref{eq:Otime}}).
}
\label{fig:Error}
\end{figure}

In order to obtain  further insight into the errors described above, we present here the 
numerical results of the multiple-quench TDNRG applied to the Anderson impurity model with a simple switching.
In Fig.~\ref{fig:Error}~(a), we switch the local level position $\varepsilon_d(t)$ by a sequence of three quenches, which 
changes the system from the mixed-valence regime to the symmetric Kondo regime.
Figure \ref{fig:Error} (b) represents the percentage deviation of the trace of 
projected density matrix from the expected value of $1$ at each time step. At the first step, the trace exactly equals $1$, but the traces 
at the later steps deviate from unity with errors less than $0.1\%$. 
Since the projected density matrix at each time step is calculated via the generalized overlap matrix elements at the same time step, 
while the latter,  $\mathcal{S}^m_{r_is_{Q_{1}}}(\tilde{\tau}_p)$, except for $\mathcal{S}^m_{r_is_{Q_{1}}}(\tilde{\tau}_0=0)$, are evaluated by
making use of the NRG approximation, we have that ${\rm Tr}[\rho^{i\to Q_{1}}(\tilde{\tau}_0)]=1$ exactly, and the trace of the projected 
density matrices at later time steps shows a finite error. 

For the continuity of the time-evolution of a local observable, we present the time-evolution of the local level occupation number $n_d$ 
in Fig.~\ref{fig:Error} (c). The gaps at the boundary of each time step, $\Delta n_d(\tilde{\tau}_p)=|n_d(\tilde{\tau}_p^+)-n_d(\tilde{\tau}_p^-)|$, 
can not be observed in the figure as they are less than $10^{-4}$. The errors in the trace of projected density matrices and the time-evolution 
of a local observable are small here due to the short time steps, and the small quench sizes. In the next section, we apply the multiple-quench TDNRG method 
to some cases, where the errors can become  more significant.

\section{Numerical results}
\label{sec:numerical}

In this section, we apply the TDNRG for multiple quenches to the Anderson impurity model for general pulses in Sec.~\ref{subsec:general-pulses}, 
clarifying the dependence of the error in the long-time limit of observables as a function of the switch-on time and pulse shape, 
for fixed initial and final states.
In Sec.~\ref{subsec:periodic-driving} we present results for periodic driving, comparing the multiple-quench TDNRG results with analytic 
continuum results in the case of the RLM and showing the applicability of the method to arbitrary finite temperatures for the non-trivial case of the 
interacting  Anderson impurity model.

\subsection{General pulses: Anderson impurity model}
\label{subsec:general-pulses}

In applying the multiple-quench TDNRG to general pulses, we use the Anderson impurity model in the strong correlation limit $U\gg \Gamma$.
We focus on switching from a given initial state (the mixed valence regime for $\epsilon_d=0$) to a given final state 
(the symmetric Kondo regime with $\varepsilon_d=-U/2$ and local level occupancy $n_{d}=1$). We shall investigate the time-evolution of the local level occupancy
for a linear ramp as a function of the switch-on time (Sec.~\ref{subsubsec:dependence-switch-on-time}), and as a function of increasingly smoother 
pulses for a fixed switch-on time (Sec.~\ref{subsubsec:dependence-pulse-shape}), comparing results in both cases to the single quench result.
We suggested in \pprI{} that the multiple-quench TDNRG may describe the long time thermodynamic limit better than the single quench case, and we
shall show below, that the results support this suggestion. Increasing the switch-on time for a given pulse, allows the system more time to 
relax to its correct thermodynamic long-time limit, which we find, while smoother pulses favor equilibration and have a similar effect. 
In brief, adiabatic changes allow for a better dissipation of energy in response to perturbations and a more accurate description of the
long-time limit.

\subsubsection{Dependence on switch-on time}
\label{subsubsec:dependence-switch-on-time}
\begin{figure}[ht]
  \centering
    \includegraphics[width=0.5\textwidth]{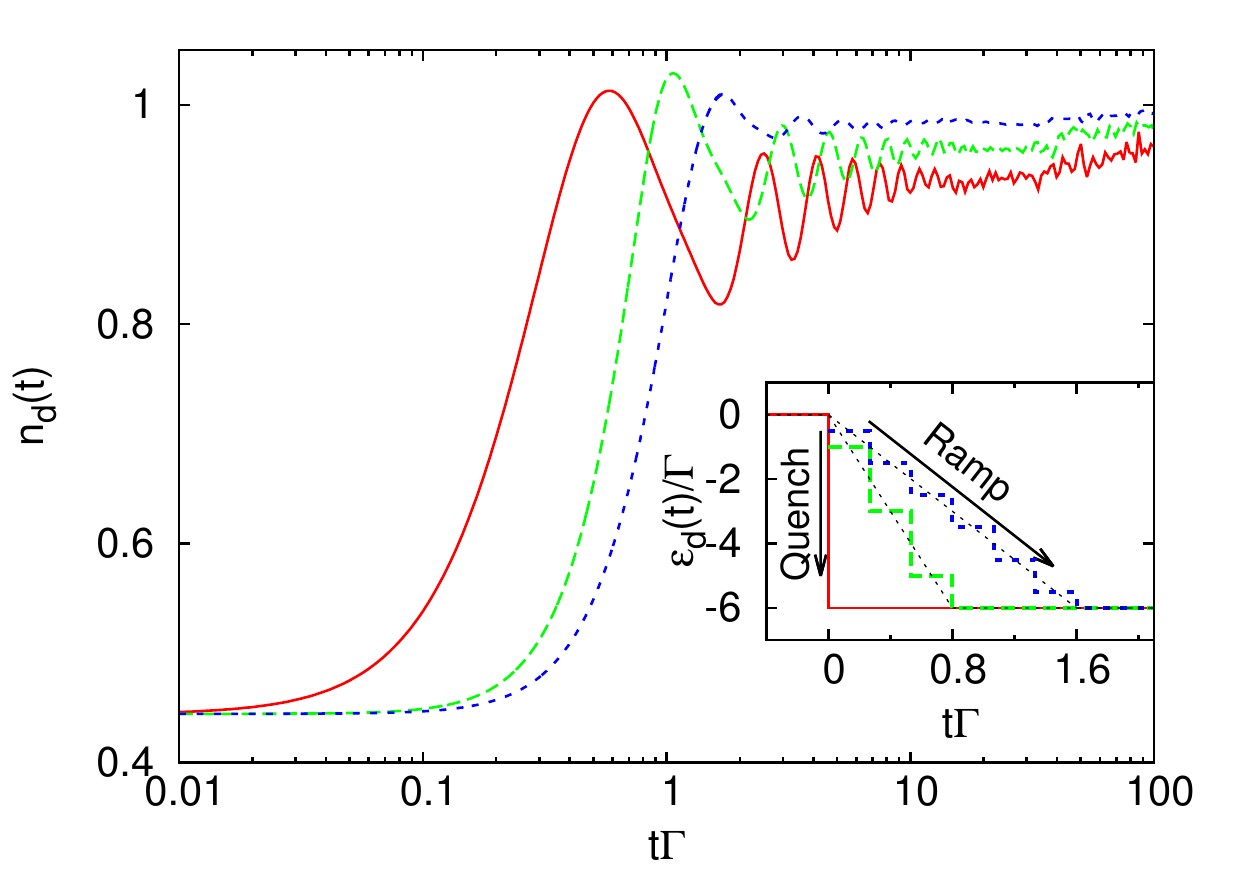}
  \caption
{
  (Color online) Time dependence of the occupation number $n_d(t)$ following pulses $\varepsilon_d(t)$ as in the inset with different switch-on times; $\tilde{\tau}_n\Gamma=0$ (single quench), $0.8$, and $1.6$ (linear ramps). In all cases, switching is from the mixed valence regime with $\varepsilon_d=0$ and $n_d\approx 0.44$ to
the symmetric Kondo regime with $\varepsilon_d=-U/2$ and $n_d=1$. The calculation is at the low temperature $T\approx 10^{-4}T_{\rm K}$. The other parameters are $U=12\Gamma$, and $\Gamma=10^{-3}D$. $T_{\rm K}\approx 2.0\times 10^{-5}D$ is the Kondo temperature in the final state. The calculations are for $\Lambda=4$, $N_z=32$, and keeping states below $E_{\rm cut}=24$.
}
\label{fig:OccVsTime}
\end{figure}

Figure \ref{fig:OccVsTime} shows the time dependence of the occupation number $n_d(t)$ upon switching the system from the mixed valence to the symmetric Kondo regime for different switch-on times: $\tilde{\tau}_n\Gamma=0$ for the sudden quench, and $\tilde{\tau}_n\Gamma=0.8$, and $1.6$ for  the linear ramps shown in the inset of Fig.~\ref{fig:OccVsTime}. The linear ramps are approximated by a sequence of smaller quenches. The number of quenches $n_{\rm quench}$ for each linear ramp is chosen by increasing it until the time-evolution converges. We use $n_{\rm quench}=20$ for the linear ramp with $\tilde{\tau}_n\Gamma=0.8$, and $n_{\rm quench}=30$ for $\tilde{\tau}_n\Gamma=1.6$. As expected, the occupation evolves in time with a delay time which increases monotonically with the switch-on time on short time scales. Note that the discontinuity in the time-evolution discussed in Sec.~\ref{sec:limits} can not be observed here, since all the gaps, $\Delta n_d(\tilde{\tau}_p)$,  at the boundaries of the time steps are less than $5\times 10^{-4}$. In the long-time limit, the linear ramp with the longer switch-on time gives us the occupation number closer to the expected value $1$, i.e., the thermodynamic value in the final state. This supports our suggestion in \pprI{} that the TDNRG can give  an improved long-time limit in the case of a sequence of small quenches over a finite time scale than in the case of a sudden large quench.

\begin{figure}[ht]
  \centering
    \includegraphics[width=0.5\textwidth]{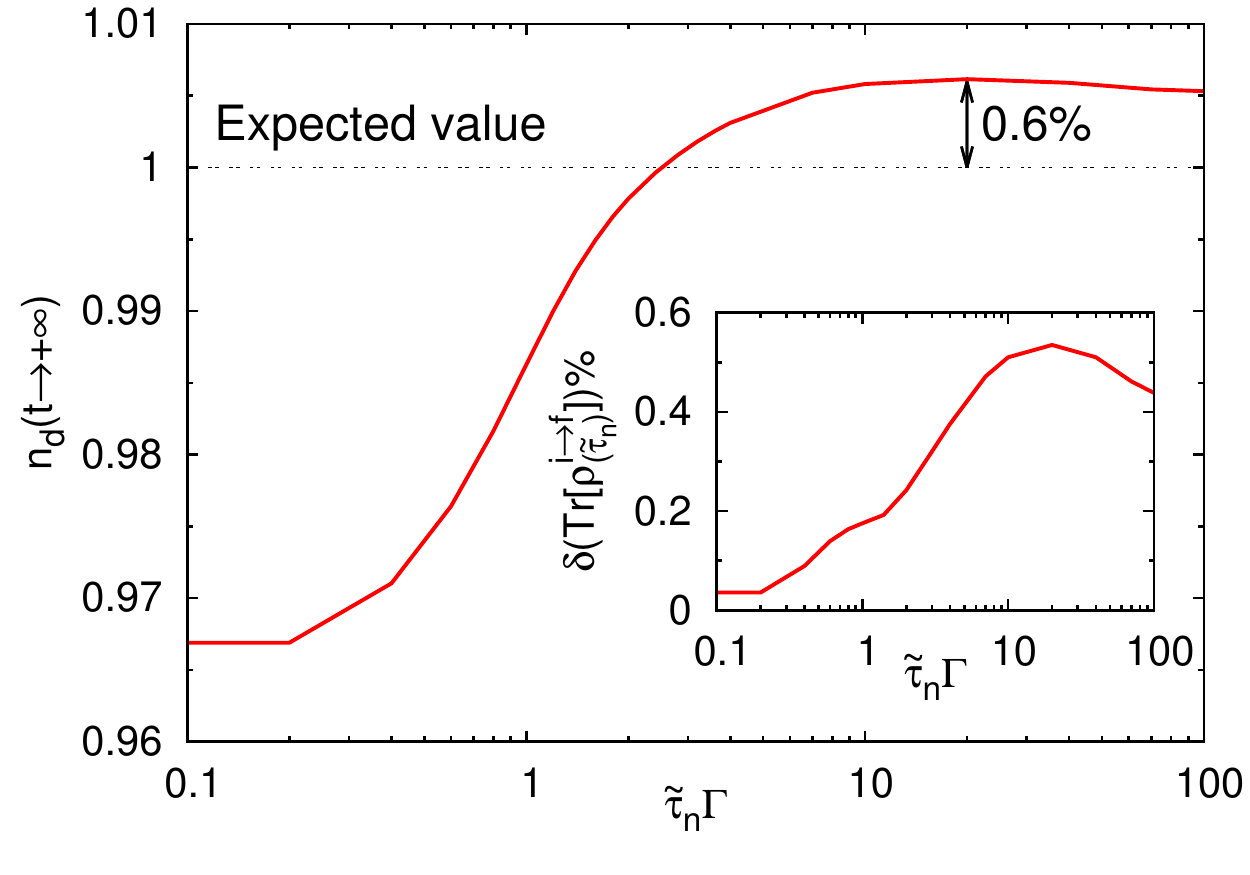}
  \caption
{
  (Color online) The occupation number in long-time limit $n_d(t\to \infty)$ vs the switch-on time $\tilde{\tau}_n$. The system is switched with a linear ramp pulse, which is approximated by a sequence of up to $100$ quenches, depending on $\tilde{\tau}_n$. The inset shows the corresponding deviation of the trace of the projected density matrix from the expected value, $1$. The calculation is for $T\approx 10^{-4}T_{\rm K}$ (essentially zero temperature). The other parameters are $U=12\Gamma$, and $\Gamma=10^{-3}D$. $T_{\rm K}\approx 2.0\times 10^{-5}D$ is the Kondo temperature in the final state. The calculations are for $\Lambda=4$, $N_z=4$, and keeping states below $E_{\rm cut}=24$.
}
\label{fig:longtimelimitVstaun}
\end{figure}

To further clarify the above discussion, we investigate the switch-on time $\tilde{\tau}_n$ dependence of the occupation number in the long-time limit, shown in Fig.~\ref{fig:longtimelimitVstaun}. 
The system here is also switched from the mixed-valence to the symmetric Kondo regime via a linear ramp, approximated as 
in the inset of Fig.~\ref{fig:OccVsTime}, but for a wider range of switch-on times $\tilde{\tau}_n$. 
For $\tilde{\tau}_n\Gamma\ge 4$, the linear ramps are approximated by a sequence of $100$ quenches, a limit set mainly by the available computer memory. 
For $\tilde{\tau}_n/\Gamma<4$, $n_{\rm quench}$ is chosen such that the occupation number in the long-time limit is converged, and we find in this case that $n_{\rm quench}<100$ suffices.
We see that the occupation number in the long-time limit approaches the expected value of $n_d=1$ as the switch-on time increases, exceeds $1$ for $\tilde{\tau}_n\Gamma \gtrsim 2$, and eventually saturates to a finite value. This finite value exceeds the expected one by $\sim 0.6\%$.

From Eq.~(\ref{eq:Otime}), we have that the occupation number in the long-time limit only depends on the diagonal elements 
of the projected density matrix at the last time step, $n_d(t\to +\infty)=\sum_{ml}\rho^{i\to f}_{ll}(m,\tilde{\tau}_n)O^m_{ll}$. 
On the other hand, from Sec.~\ref{sec:limits} we learn that the TDNRG calculation for multiple quenches gives rise to an error in 
the trace of the projected density matrix, ${\rm Tr}[\rho^{i\to f}(\tilde{\tau}_{n})]=\sum_{ml}\rho^{i\to f}_{ll}(m,\tilde{\tau}_n)$. 
One may, therefore, raise a question concerning the switch-on time dependence of the occupation number in the long-time limit, namely, 
whether the occupation number in the long-time limit is really getting closer to the expected value with increasing switch-on times, or, whether 
this is a result of the accumulated error of the projected density matrix, e.g., as shown in Fig.~\ref{fig:Error}~(b). 
To clarify this, we show the error in the trace of the corresponding projected density matrix at the last time step versus the switch-on time 
in the inset to Fig.~\ref{fig:longtimelimitVstaun}. 
This error is seen to also increase with increasing switch-on time, but does not exceed $0.6\%$, and also starts to saturate at longer 
switch-on times. This suggests that the error in the projected density matrix results in the small $0.6\%$ deviation of $n_d(t\to \infty)$ 
from its expected long-time limit of $1$ observed in Fig.~\ref{fig:longtimelimitVstaun}. 
Therefore, we conclude that longer switch-on times really result in the occupation number coming closer to its expected value in the 
long-time limit. Since the source of the error in the trace ${\rm Tr}[\rho^{i\to f}(\tilde{\tau}_{n})]$ stems from the NRG approximation used in the 
evaluation of the generalized overlap matrix elements (see Sec.~\ref{sec:limits}), improved schemes for evaluating the latter may allow the
multiple-quench formalism to obtain the long-time limit of observables exactly. 

\subsubsection{Dependence on pulse shape}
\label{subsubsec:dependence-pulse-shape}
\begin{figure}[ht]
  \centering
    \includegraphics[width=0.5\textwidth]{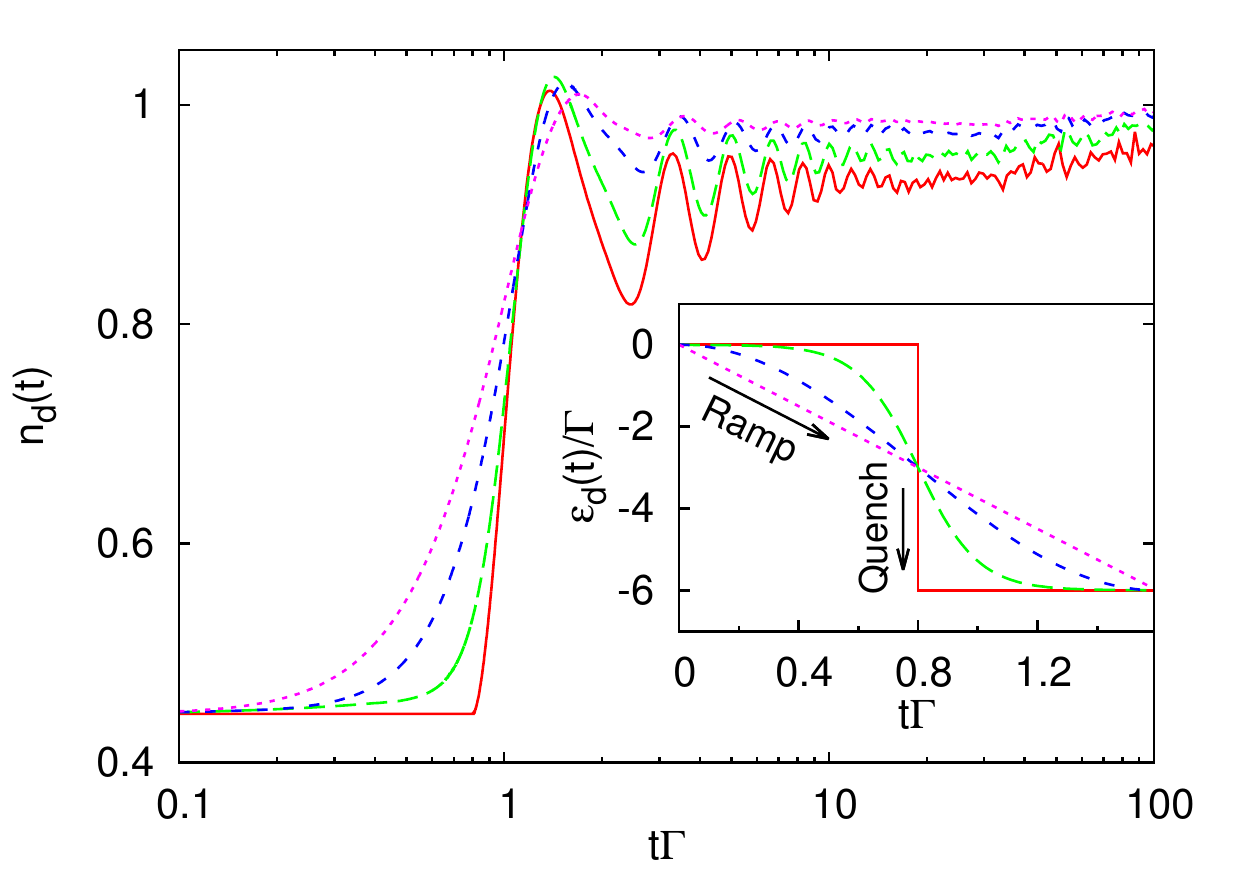}
  \caption
{
  (Color online) Time dependence of the occupation number $n_d(t)$ in response to pulses with different shapes but with a constant switch-on time $\tilde{\tau}_n=1.6/\Gamma$; $\varepsilon_d(t)$ is shown in the inset for the four pulse shapes [square (solid line), logistic (long-dashed line), trigonometric (dashed line), and linear (dotted line)]. The calculation is at the low temperature $T\approx 10^{-4}T_{\rm K}$. The other parameters are $U=12\Gamma$, and $\Gamma=10^{-3}D$. $T_{\rm K}\approx 2.0\times 10^{-5}D$ is the Kondo temperature in the final state. The calculations are for $\Lambda=4$, $N_z=32$, and keeping states below $E_{\rm cut}=24$.
}
\label{fig:OccVsTimePulse}
\end{figure}

We now turn to the effect of the pulse shape on the time-evolution for a fixed switch-on time. Figure~\ref{fig:OccVsTimePulse} shows the time dependence of the occupation number upon switching the system from a fixed initial state in the mixed valence regime (that for $\varepsilon_d=0$) to the final state defined by 
the symmetric Kondo regime ($\varepsilon_d=-U/2$) using pulses of different shape while maintaining a fixed switch-on time: the step function (the sudden quench), the logistic, trigonometric, and linear functions (smooth pulses) as represented in the inset to Figure~\ref{fig:OccVsTimePulse}. For comparison between the cases of a sudden quench and general smooth pulses, we shift the sudden quench to start at $\tilde{\tau}_0\Gamma=0.8$, and set the switch-on time equal for all three smooth pulses ($\tilde{\tau}_n\Gamma=1.6$) with $\tilde{\tau}_0\Gamma=0$. Each pulse is approximated by a sequence of $30$ quenches, described by the same set of $\{H^{Q_0},\dots,H^{Q_{n+1}}\}$. Due to the different pulse shapes, however, we have different sets of $\{\tilde{\tau}_1,\dots,\tilde{\tau}_{n}\}$ for each pulse. As expected, the time-evolution of the occupation number at short times is more rapid for pulses which vary more rapidly. 
As in Fig.~\ref{fig:OccVsTime}, the discontinuity in the time-evolution can not be observed here since the gaps at the boundaries of the time steps are less than $5\times 10^{-4}$. In the long-time limit, the smoother varying pulses result in occupation numbers closer to the expected value of $n_d=1$. Together with the switch-on time dependence of the occupation number in the long-time limit in Fig.~\ref{fig:longtimelimitVstaun}, this suggests that the  TDNRG for general pulses gives improved results the smoother the pulse. Smoother pulses also imply increased adiabaticity favoring energy dissipation and relaxation to the correct long-time limit.
Note also, the gradual disappearance of the oscillations with increasing switch-on time in Fig.~\ref{fig:OccVsTime}, and with increasing 
smoothness of the pulse in Fig.~\ref{fig:OccVsTimePulse}. The former trend has been noted before in the context of the interacting resonant 
level model. \cite{Kennes2012b} In general, this suppression of ringing correlates  
with increased adiabaticity.

\subsection{Periodic driving}
\label{subsec:periodic-driving}
In this section we first apply the multiple-quench TDNRG to periodic driving in the exactly solvable RLM and compare the numerical 
results with analytic continuum results. We consider square and triangular periodic driving of the local level position (Sec.~\ref{subsubsec:rlm}). 
We next apply the TDNRG to the non-trivial case of the 
strongly correlated Anderson impurity model $U\gg \Gamma$ with a triangular periodic driving of the local level, showing, 
in particular results for the time-evolution of the
occupation number at arbitrary finite temperatures (Sec.~\ref{subsubsec:temperature-dependence}).

\subsubsection{RLM: Comparison with exact results}
\label{subsubsec:rlm}
In the application of the multiple-quench TDNRG method to the RLM, we shall further check the accuracy of the method by comparing the time-evolution of the occupation number to the analytical results in the wide-band limit. The analytic expression for the occupation number following a single quench\cite{Anders2006} is generalized to the multiple-quench case as follows,
\begin{align}
&n_d(\tilde{\tau}_{p+1}>t\ge \tilde{\tau}_p)=\rho_F\int_{-\infty}^{+\infty}f(\varepsilon)|A(\varepsilon,t)|^2 d\varepsilon,\label{eq:analytic_nd}\\
&A(\varepsilon,\tilde{\tau}_{p+1}>t\ge \tilde{\tau}_p)=\frac{V_{Q_{p+1}}e^{-i\varepsilon t}}{i(\varepsilon^{Q_{p+1}}_{d}-\varepsilon)+\Gamma_{Q_{p+1}}}\label{eq:analytic_A}\\
&-e^{-i(\varepsilon^{Q_{p+1}}_{d}+\Gamma_{Q_{p+1}}) (t-\tilde{\tau}_p)}\left(\frac{V_{Q_{p+1}}e^{-i\varepsilon \tilde{\tau}_p}}{i(\varepsilon^{Q_{p+1}}_{d}-\varepsilon)+\Gamma_{Q_{p+1}}}-A(\varepsilon,\tilde{\tau}_p)\right)\nonumber,
\end{align}
in which $A(\varepsilon,\tilde{\tau}_{p+1}>t\ge \tilde{\tau}_p)$ is calculated recursively with $\displaystyle A(\varepsilon,\tilde{\tau}_0)=\frac{V_{i}}{i(\varepsilon^{i}_{d}-\varepsilon)+\Gamma_{i}}$ corresponding to the initial state. $\rho_F$ is the density of state of the fermionic bath, $f(\varepsilon)$ is the Fermi distribution, $\Gamma_{Q_{p}}=\pi\rho_F|V_{Q_{p}}|^2$, and $\{V_{Q_{p}}, \varepsilon_d^{Q_{p}}\}$ are the hybridization and local level associated with the quench Hamiltonian $H^{Q_{p}}$. In both the analytic and the TDNRG calculations, we  approximate a smooth pulse, or here, a train of pulses, by exactly the same sequence of small quenches. Thus, we can compare directly the exact continuum results with those of the TDNRG approach.

\begin{figure}[ht]
  \centering
    \includegraphics[width=0.5\textwidth]{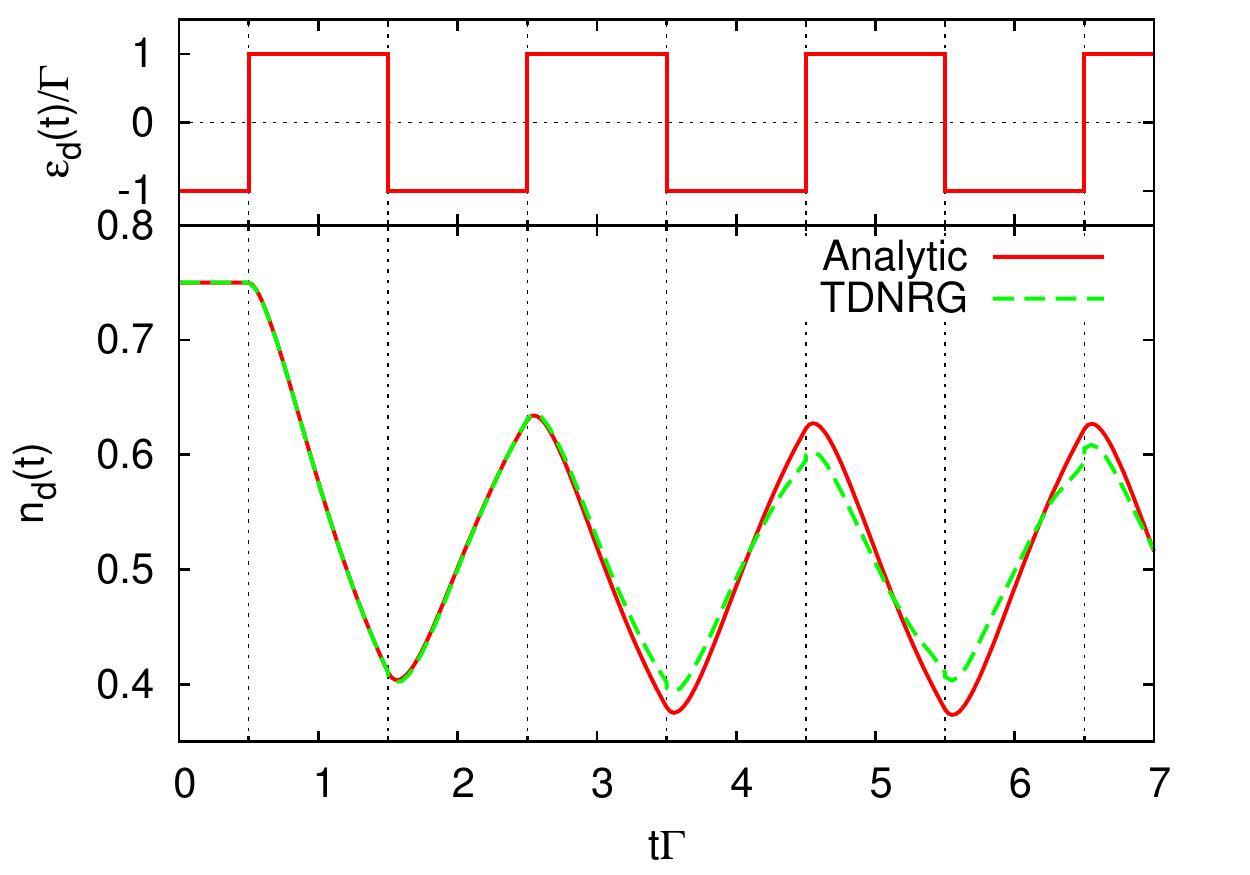}
  \caption
{
  (Color online) Application of the multiple-quench TDNRG to the RLM with the square periodic switching $\varepsilon_d(t)$ as in the upper panel figure. The lower panel shows the time-evolution of the occupation number at the low temperature $T\approx 10^{-4}\Gamma$. The other parameters are $U=0\Gamma$, and $\Gamma=10^{-3}D$. The calculations are for $\Lambda=4$, $z$ averaging with $N_z=16$, and keeping states below $E_{\rm cut}=24$.
}
\label{fig:periodicSquare}
\end{figure}

In Fig.~\ref{fig:periodicSquare} (lower panel), we show the time-evolution of the occupation number, following the periodic switching, represented in the upper panel of Fig.~\ref{fig:periodicSquare}. One sees that the occupation numbers calculated with the TDNRG and the analytical Eqs.~(\ref{eq:analytic_nd}-\ref{eq:analytic_A}) both oscillate in time with the same frequency as the driving. The two results agree very well with each other up to $t{\Gamma}<3$ and deviate for longer times. The discontinuity in the time-evolution of the occupation number in the TDNRG calculation can be observed here with visible gaps $\Delta n_d(t)$ at the boundaries of the time steps, $t{\Gamma}=4.5,5.5$, and $6.5$, while the analytical result is obviously continuous. The difference between the results of the two calculations comes partly from the fact that the TDNRG calculation is based on the logarithmic discretization of the conduction band, while the analytic calculation is carried out in the continuum limit. However, the NRG approximation also contributes to  this difference, resulting in the  observed discontinuities which increase in size with increasing time. We expect, in general, following the discussion of pulse shapes on cumulative errors in Sec.~\ref{subsec:general-pulses}, that smoother driving will show reduced errors at longer times, a topic we discuss next.

\begin{figure}[ht]
  \centering
    \includegraphics[width=0.5\textwidth]{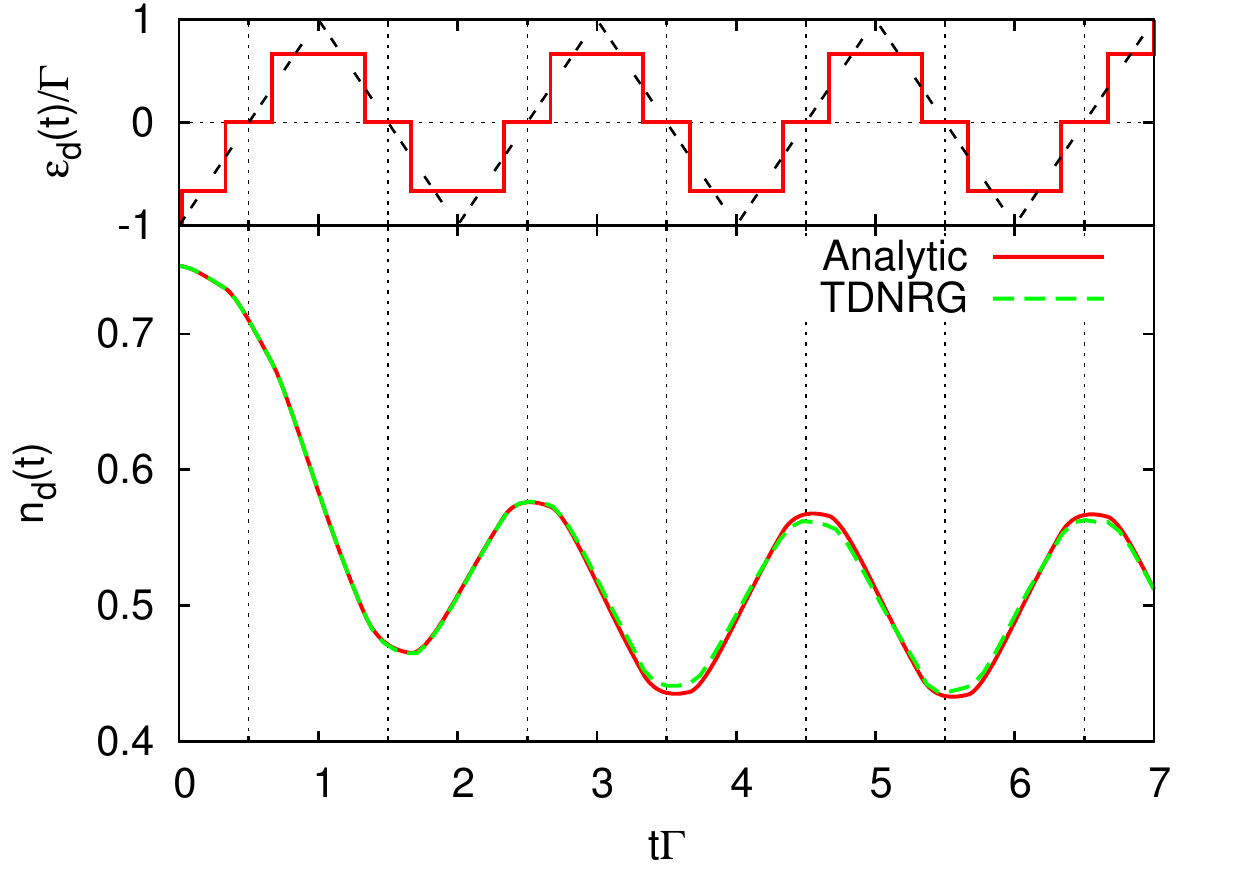}
  \caption
{
  (Color online) As in Fig.~\ref{fig:periodicSquare}, but with triangular periodic switching approximated by the  sequence of small quenches shown in the upper panel. The lower panel shows the time-evolution of the occupation number $n_d(t)$ for this case and the parameters are as in Fig.~\ref{fig:periodicSquare}.
}
\label{fig:periodicTriangle}
\end{figure}

Figure~\ref{fig:periodicTriangle} shows the time-evolution of the occupation number (lower panel) in response to a triangular periodic driving, which is approximately replaced by a sequence of small quenches, represented in the upper panel of Fig.~\ref{fig:periodicTriangle}. We set the square and triangular periodic drivings in Figs.~\ref{fig:periodicSquare} and \ref{fig:periodicTriangle} to have the same frequency and phase, therefore the oscillations of the occupation numbers calculated by either the TDNRG or the analytical expression in these two figures are period- and phase-matching. In contrast, the amplitude of the oscillations in the occupation numbers in the two cases differ, with triangular switching resulting in a smaller amplitude. In the TDNRG calculations for the square and triangular drivings, we have used the same discretization parameter $\Lambda=4$. However, in comparison to the case of square switching in Figs.~\ref{fig:periodicSquare}, we see that the TDNRG result for the time-evolution of the occupation number with triangular periodic driving exhibits better agreement to the analytical result, and less significant gaps at the boundaries of the time steps. This suggests that the TDNRG calculation for multiple quenches gives the time-evolution of a local observable in closer agreement to the 
exact result if each quench size is small enough and for sufficiently smooth trains of pulses.

\subsubsection{Periodically driven Anderson model: temperature dependence}
\label{subsubsec:temperature-dependence}
\begin{figure}[ht]
  \centering
    \includegraphics[width=0.5\textwidth]{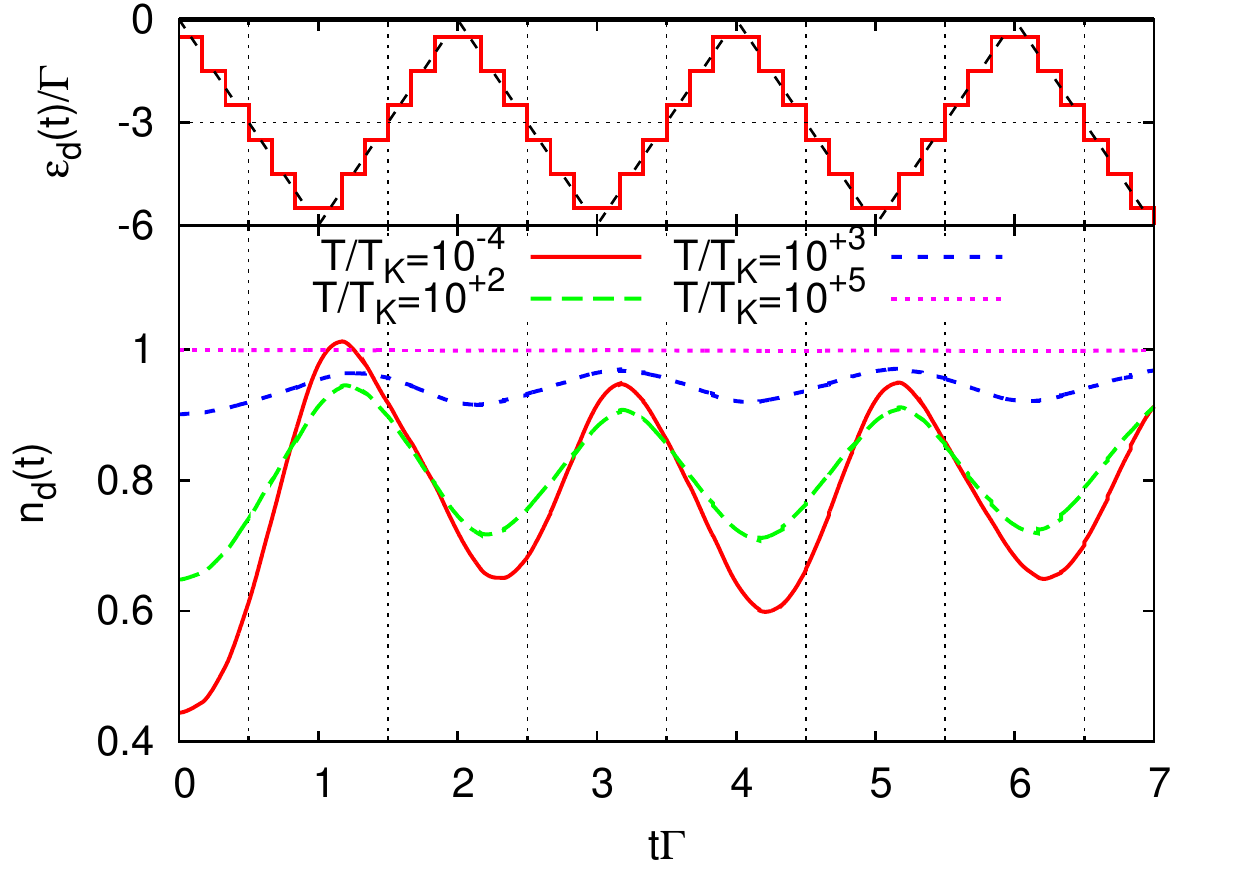}
  \caption
{
  (Color online) Application to the Anderson impurity model with triangular periodic switching $\varepsilon_d(t)$ as shown in the upper panel. The lower panel shows the time-evolution of the occupation number for a number of temperatures $T/T_{\rm K}$, ranging from very low, $10^{-4}$ (essentially zero temperature), to very high, $10^{+5}$ (comparable to band width). The other parameters are $U=12\Gamma$, and $\Gamma=10^{-3}D$. $T_{\rm K}\approx 2.0\times 10^{-5}D$ is the Kondo temperature in the final state. The calculations are for $\Lambda=4$, $N_z=16$, and keeping states below $E_{\rm cut}=24$.
}
\label{fig:AIMperiodicTriangle}
\end{figure}

So far we have only shown results for very low (essentially zero) temperature. However, the multiple-quench TDNRG formalism is also 
applicable to arbitrary finite temperatures since it is based on the FDM approach. Temperature effects are particularly important for 
interacting systems, such as the Kondo or Anderson impurity models. To illustrate the applicability of our formalism to finite temperature,
we show in Fig.~\ref{fig:AIMperiodicTriangle} the time dependence of the occupation number in the Anderson impurity model 
upon switching the system periodically between the mixed valence and the symmetric Kondo regime at four different temperatures, $T/T_{\rm K}=10^{-4},10^{+2},10^{+3}$, and $10^{+5}$. We use triangular switching, and approximately replace this by a sequence of quenches, $n_{\rm quench}=44$, as shown in the upper part of Fig.~\ref{fig:AIMperiodicTriangle}. At each temperature, one can see that the occupation number oscillates in time with the driving frequency. The oscillation amplitude decreases with increasing temperature, and, eventually, at the very highest temperature $T=10^{5}T_{\rm K}\approx 2000\Gamma \gg \Gamma$ (comparable to the bandwidth), the oscillations disappear. 
In this limit, where $T\gg |\varepsilon_d(t)|,U, \Gamma$, all four states of the impurity are equally occupied and the average local level occupancy 
acquires the time-independent value of $n_{d}=1$.
As in the application of the multiple-quench TDNRG to the RLM with triangular periodic driving, the gaps at the boundaries of the time steps are sufficiently small as to not be observable in the figure at all temperatures.

\section{Conclusions}
\label{sec:summary}
In this paper, we numerically implemented the TDNRG formalism for multiple quenches, derived in our previous paper\cite{Nghiem2014}, to study the response of a quantum impurity system to general pulses and periodic switching. Several limiting cases allowed us to test the correctness and accuracy of our numerical implementation.

For general pulses, with switching between a given initial and a given final state, we applied the method to the Anderson impurity model and investigated how the error in the long-time limit depends on the switch-on time and the pulse shape. 
We found that the long-time limit becomes more accurate with increasing switch-on times for a given pulse shape (a linear ramp) and with increasing smoothness of the pulse 
(for a fixed switch-on time). We interpret this as implying that longer switch-on times or smoother pulses, i.e., increased adiabaticity, favor equilibration of the system to its correct long-time limit.
The switch-on time and pulse shape dependence of the long-time limit supports our suggestion in the previous paper that the long-time limit can be improved if the system is 
switched by a sequence of many small quenches over a finite time scale instead of a sudden large quench. The multiple-quench TDNRG formalism 
therefore provides an algorithmic improvement in obtaining the long-time limit of observables, as compared to the single-quench formalism in \pprI{}. 
Nevertheless, as discussed in Ref.~\onlinecite{Rosch2012}, the use in NRG calculations of a Wilson chain, which has only  a finite (non-extensive) heat capacity, may 
prohibit thermalization of local observables to their exact thermodynamic values at long times. Support for this comes from the observation in \pprI{} that 
the long-time limit of observables is indeed improved for $\Lambda\to 1^{+}$, in which a Wilson chain ($\Lambda>1$) approaches a continuum bath ($\Lambda=1$). 
Since the limit $\Lambda\to 1^+$ is impractical in NRG calculations \cite{Wilson1975}, it would be interesting in the future to explore ways of including 
a coupling to a thermal reservoir within NRG in order to address the above problem.

For periodic driving, we compared the multiple-quench TDNRG calculations to available exact analytic results for the RLM. 
In the short to intermediate time range, the time-evolution of the occupation number shows better agreement to the analytic results in the case of 
triangular periodic switching than in the case of square periodic switching. This also suggests that the time-evolution is more accurate for smoother periodic pulses
than for less smooth periodic pulses (e.g., square pulses).
Finally, we applied the TDNRG to the Anderson impurity model with periodic driving and demonstrated the validity of the formalism to an arbitrary finite temperature. 

We identified a source of error in the multiple-quench TDNRG, absent in the single-quench case, which
is due to the use of the NRG approximation in the time-evolution factors entering the generalized overlap matrix elements (Sec.~\ref{subsec:formalism} and Sec.~\ref{sec:limits}). While the errors are small in many situations, see Sec.~\ref{sec:limits} and \ref{sec:numerical}, they can become significant after many cycles in the case of periodic driving. It would therefore be of interest in the future to devise
alternative schemes for evaluating the generalized overlap matrix elements, in order to reduce or overcome this source of error.

In future, it would be interesting to apply the present formalism to pump-probe spectroscopies of magnetic adatoms to calculate lifetimes of excited states,
\cite{Loth2010} to quantum dots in time-dependent fields, \cite{Medvedyeva2013, Nordlander1999,Plihal2000,Bruder1994,Kogan2004,Haupt2013} and, with a suitable generalization to spectral functions,  to time-resolved photoemission and related spectroscopies. \cite{Perfetti2006,Loth2010,Iyoda2014,Freericks2009,Eckstein2008} The latter generalization would also be of interest in the context of non-equilibrium dynamical mean field theory.\cite{Freericks2006,Aoki2014}

\begin{acknowledgments}
We thank A. Weichselbaum for useful comments and acknowledge supercomputer support by the John von Neumann institute for Computing (J\"ulich).
\end{acknowledgments}
\appendix
\begin{widetext}
\section{Trace of the Projected density matrix}
\label{sec:trace-projected-density-matrix}
For the proof of the conservation of the trace of the projected density matrices, we have, starting from the equation preceding Eq.~(\ref{eq:rhomultiquenches}), 
\begin{align}
&{\rm Tr}[\rho^{i\to Q_{p+1}}(\tilde{\tau}_{p})]
=\sum_{mle}{_{Q_{p+1}}}\langle lem|e^{-iH^{Q_p}{\tau}_p}\dots e^{-iH^{Q_1}{\tau}_1}\rho e^{iH^{Q_1}{\tau}_1}\dots e^{iH^{Q_p}{\tau}_p}| lem\rangle{_{Q_{p+1}}}\nonumber\\
=&\sum_{mle}\sum_{m_1l_1e_1}{_{Q_{p+1}}}\langle lem|e^{-iH^{Q_p}{\tau}_p}\dots e^{-iH^{Q_1}{\tau}_1}|l_1e_1m_1\rangle{_i}w_{m_1} \frac{e^{-\beta E_{l_1}^{m_1}}}{\tilde{Z}_{m_1}}{_i}\langle l_1e_1m_1|e^{iH^{Q_1}{\tau}_1}\dots e^{iH^{Q_p}{\tau}_p}| lem\rangle_{Q_{p+1}}\nonumber\\
=&\sum_{mle}\sum_{m_1l_1e_1}{_i}\langle l_1e_1m_1|e^{iH^{Q_1}{\tau}_1}\dots e^{iH^{Q_p}{\tau}_p}| lem\rangle_{Q_{p+1}}{_{Q_{p+1}}}\langle lem|e^{-iH^{Q_p}{\tau}_p}\dots e^{-iH^{Q_1}{\tau}_1}|l_1e_1m_1\rangle{_i}w_{m_1} \frac{e^{-\beta E_{l_1}^{m_1}}}{\tilde{Z}_{m_1}}\nonumber\\
=&\sum_{m_1l_1e_1}{_i}\langle l_1e_1m_1|e^{iH^{Q_1}{\tau}_1}\dots e^{iH^{Q_p}{\tau}_p}e^{-iH^{Q_p}{\tau}_p}\dots e^{-iH^{Q_1}{\tau}_1}|l_1e_1m_1\rangle{_i}
w_{m_1} \frac{e^{-\beta E_{l_1}^{m_1}}}{\tilde{Z}_{m_1}}\nonumber\\
=&\sum_{m_1l_1}w_{m_1} \frac{e^{-\beta E_{l_1}^{m_1}}}{{Z}_{m_1}}=1.
\end{align}

This is conserved at any step, so that we have ${\rm Tr}[\rho^{i\to Q_{p+1}}(\tilde{\tau}_{p})]=\dots={\rm Tr}[\rho^{i\to Q_{1}}(\tilde{\tau}_{0})]=1$. 

Notice, that in this proof, 
we have not made use of the NRG approximation for the generalized overlap matrix elements defined in Eq.~(\ref{eq:overlapTime}) that appear in the above expression for the
trace of the projected density matrices. In practice, however, the NRG approximation is required to obtain feasible expressions for these matrix elements, such as the 
recursive expressions  in Eqs.~(\ref{eq:totalStau}-\ref{eq:totalStau--}) for each time step (except for $p=1$ where no NRG approximation enters these matrix elements). 
Use of the latter in the above expression for the trace results, then, in a finite error in all but the first projected density matrix.

\section{Continuity of the time-evolution of a local observable}
\label{sec:continuity}
For the continuity, we start from the general equation [see Eq.~(\ref{eq:Otime})]
\begin{align}
O(t\ge \tilde{\tau}_p)=&\sum_{mrse}^{\notin KK'} {_{Q_{p+1}}}\langle rem|e^{-iH^{Q_p}{\tau}_p}\dots e^{-iH^{Q_1}{\tau}_1}\rho e^{iH^{Q_1}{\tau}_1}\dots e^{iH^{Q_p}{\tau}_p}| sem\rangle{_{Q_{p+1}}}e^{-i(E^m_r-E^m_s)(t-\tilde{\tau}_p)}{_{Q_{p+1}}}\langle sem|\hat{O}| rem\rangle_{Q_{p+1}}\label{eq:general},
\end{align}
\begin{align}
\text{then}\quad O(t\to\tilde{\tau}^-_p)=&\sum_{mrse}^{\notin KK'} {_{Q_p}}\langle rem|e^{-iH^{Q_{p-1}}{\tau}_{p-1}}\dots e^{-iH^{Q_1}{\tau}_1}\rho e^{iH^{Q_1}{\tau}_1}\dots e^{iH^{Q_{p-1}}{\tau}_{p-1}}| sem\rangle{_{Q_{p}}} e^{-i(E^m_r-E^m_s)(\tilde{\tau}_p-\tilde{\tau}_{p-1})}{_{Q_{p}}}\langle sem|\hat{O}| rem\rangle_{Q_{p}}\nonumber \\
=&\sum_{mrse}^{\notin KK'} {_{Q_p}}\langle rem|e^{-iH^{Q_{p-1}}{\tau}_{p-1}}\dots e^{-iH^{Q_1}{\tau}_1}\rho e^{iH^{Q_1}{\tau}_1}\dots e^{iH^{Q_{p-1}}{\tau}_{p-1}}| sem\rangle{_{Q_{p}}} e^{-i(E^m_r-E^m_s){\tau}_p}{_{Q_{p}}}\langle sem|\hat{O}| rem\rangle_{Q_{p}}\label{eq:Otaup-},
\end{align} 
\begin{align}
&\text{and}\quad O(t\to\tilde{\tau}^+_p)\nonumber\\
=&\sum_{mrse}^{\notin KK'} {_{Q_{p+1}}}\langle rem|e^{-iH^{Q_p}{\tau}_p}\dots e^{-iH^{Q_1}{\tau}_1}\rho e^{iH^{Q_1}{\tau}_1}\dots e^{iH^{Q_p}{\tau}_p}| sem\rangle{_{Q_{p+1}}}e^{-i(E^m_r-E^m_s)(\tilde{\tau}_p-\tilde{\tau}_p)}{_{Q_{p+1}}}\langle sem|\hat{O}| rem\rangle_{Q_{p+1}}\nonumber\\
=&\sum_{mrse}^{\notin KK'} {_{Q_{p+1}}}\langle rem|e^{-iH^{Q_p}{\tau}_p}\dots e^{-iH^{Q_1}{\tau}_1}\rho e^{iH^{Q_1}{\tau}_1}\dots e^{iH^{Q_p}{\tau}_p}| sem\rangle{_{Q_{p+1}}} {_{Q_{p+1}}}\langle sem|\hat{O}| rem\rangle_{Q_{p+1}}\nonumber\\
=&\sum_{mrse}^{\notin KK'}\sum_{m_1l_1e_1}\sum_{m_2l_2e_2} {_{Q_{p+1}}}\langle rem|l_1e_1m_1\rangle{_{Q_p}} {_{Q_p}}\langle l_1e_1m_1|e^{-iH^{Q_p}{\tau}_p}\dots e^{-iH^{Q_1}{\tau}_1}\rho  e^{iH^{Q_1}{\tau}_1}\dots e^{iH^{Q_p}{\tau}_p}| l_2e_2m_2\rangle{_{Q_p}}{_{Q_p}}\langle l_2e_2m_2|sem\rangle{_{Q_{p+1}}} {_{Q_{p+1}}}\langle sem|\hat{O}| rem\rangle_{Q_{p+1}}\nonumber\\
=&\sum_{mrse}^{\notin KK'}\sum_{m_1l_1e_1}\sum_{m_2l_2e_2}  {_{Q_p}}\langle l_1e_1m_1|e^{-iH^{Q_p}{\tau}_p}\dots e^{-iH^{Q_1}{\tau}_1}\rho  e^{iH^{Q_1}{\tau}_1}\dots e^{iH^{Q_p}{\tau}_p}| l_2e_2m_2\rangle{_{Q_p}} {_{Q_p}}\langle l_2e_2m_2|sem\rangle{_{Q_{p+1}}} {_{Q_{p+1}}}\langle sem|\hat{O}| rem\rangle_{Q_{p+1}}{_{Q_{p+1}}}\langle rem|l_1e_1m_1\rangle{_{Q_p}}
\label{eq:Otaup+}.
\end{align}
We can prove that $\sum_{mrse}^{\notin KK'}{_{Q_p}}\langle l_2e_2m_2|sem\rangle{_{Q_{p+1}}} {_{Q_{p+1}}}\langle sem|\hat{O}| rem\rangle_{Q_{p+1}}{_{Q_{p+1}}}\langle rem|l_1e_1m_1\rangle{_{Q_p}}={_{Q_p}}\langle l_2e_2m_2|\hat{O}| l_1e_1m_1\rangle{_{Q_p}}$\cite{Nghiem2014}. Substituting this into Eq.~(\ref{eq:Otaup+}), we have that
\begin{align}
O(t\to\tilde{\tau}^+_p)=&\sum_{m_1l_1e_1}\sum_{m_2l_2e_2}  {_{Q_p}}\langle l_1e_1m_1|e^{-iH^{Q_p}{\tau}_p}\dots e^{-iH^{Q_1}{\tau}_1}\rho  e^{iH^{Q_1}{\tau}_1}\dots e^{iH^{Q_p}{\tau}_p}| l_2e_2m_2\rangle{_{Q_p}} {_{Q_p}}\langle l_2e_2m_2|\hat{O}| l_1e_1m_1\rangle{_{Q_p}}\nonumber\\
=&\sum_{m_1rse_1}^{\notin KK'} {_{Q_p}}\langle re_1m_1|e^{-iH^{Q_p}{\tau}_p}\dots e^{-iH^{Q_1}{\tau}_1}\rho  e^{iH^{Q_1}{\tau}_1}\dots e^{iH^{Q_p}{\tau}_p}| se_1m_1\rangle{_{Q_p}}{_{Q_p}}\langle se_1m_1|\hat{O}| re_1m_1\rangle{_{Q_p}}\nonumber\\
=&\sum_{m_1rse_1}^{\notin KK'} {_{Q_p}}\langle re_1m_1|e^{-iH^{Q_{p-1}}{\tau}_{p-1}}\dots e^{-iH^{Q_1}{\tau}_1}\rho  e^{iH^{Q_1}{\tau}_1}\dots e^{iH^{Q_{p-1}}{\tau}_{p-1}}| se_1m_1\rangle{_{Q_p}}e^{i(E^{m_1}_s-E^{m_1}_r){\tau}_p}{_{Q_p}}\langle se_1m_1|\hat{O}| re_1m_1\rangle{_{Q_p}}\label{eq:Otaup+2}.
\end{align}
Clearly, from equations (\ref{eq:Otaup-}) and (\ref{eq:Otaup+2}), we have that $O(t\to\tilde{\tau}^+_p)=O(t\to\tilde{\tau}^-_p)$. 

As in appendix~\ref{sec:trace-projected-density-matrix}, the above proof of the continuity of observables uses the general form for the generalized overlap matrix elements 
[Eq.~(\ref{eq:overlapTime})]. Once these are reduced to their recursive form in Eqs.~(\ref{eq:totalStau}-\ref{eq:totalStau--}) via the use of the NRG approximation, 
continuity is only guaranteed for $t\to 0^{+}$, i.e., for the short-time limit of observables, as in \pprI{}.
\end{widetext}

\bibliography{noneq-nrg}
\end{document}